# Thermal Activation and Quantum Field Emission in a Sketch-Based Oxide Nano Transistor

Cheng Cen[1], Daniela F. Bogorin[1], Jeremy Levy[1*]

*Corresponding author, jlevy@pitt.edu

[1]Department of Physics and Astronomy, University of Pittsburgh, Pittsburgh, PA 15260, USA

**The interface between polar $LaAlO_3$ and non-polar $SrTiO_3$ exhibits a remarkable variety of electronic behavior[1] associated with the formation of an interfacial quasi-two-dimensional electron gas (q-2DEG)[2-8]. By "sketching" patterns of charge on the top $LaAlO_3$ surface, the $LaAlO_3/SrTiO_3$ interface conductance can be controlled with near-atomic spatial resolution[9]. Using this technique, a sketch-based oxide nanotransistor (SketchFET) was demonstrated[10] with a minimum feature size of just two nanometers. Here we report direct measurements of the potential barriers and electronic coupling between nanowire segments within a SketchFET device. Near room temperature, switching is governed by thermally activated field emission from the nanowire gate. Below *T*=150 K, a crossover to quantum field emission is observed that is sensitive to structural phase transitions in the $SrTiO_3$ layer. This direct measurement of the source-drain and gate-drain energy barriers is crucial for the development of room-temperature logic and memory elements and low-temperature quantum devices.**

The heterostructure consisting of 3 unit cells (uc) of $LaAlO_3$ grown on $TiO_2$-terminated $SrTiO_3$ (3uc-LAO/STO) can be regarded as an ultra-dense two-dimensional network of floating-gate field-effect transistors. A positive bias applied to a conductive atomic-force microscope (c-AFM) tip surface charges the top $LaAlO_3$ surface[10,11], locally switching the interface to a conducting state; a negative applied bias discharges the top $LaAlO_3$ surface, locally restoring the insulating state. The nanoscale writing and erasing can be regarded as a form of reversible modulation doping, using donors that are placed approximately one nanometer from the interface. A variety of structures and devices have been demonstrated at room temperature[9,10,12].

A central question arises: what is the depth of the electronic confinement potential in these devices? A reliable method for determining these energy barriers is through temperature-dependent transport measurements. A convenient structure for quantifying these energy barriers is a SketchFET itself, since the source-drain barrier can be tuned by the gate electrode. Furthermore, the energy barriers that are measured can be directly related to the performance of the device.

SketchFETs are created from an oxide heterostructure consisting of 3.3 unit cells of $LaAlO_3$ grown on $TiO_2$-terminated $SrTiO_3$ substrates. The films were grown at the University of Augsburg by pulsed laser deposition using parameters that are described elsewhere[9,10]. SketchFET structures are written at the interface between the two oxides using a c-AFM probe (Fig. 1a). Two SketchFET devices are created under nominally identical conditions with qualitatively reproduced characteristics observed. For consistency, results are shown for only one of them. The particular SketchFET device discussed here (Fig. 1b) consists of three nanowire sections: "source", "drain" and "gate". Each nanowire is written with a tip bias $V_{tip}$= +10V, which produces an approximately 16 nm wire width (Fig S1). A comparably-sized potential barrier in the channel between the source and drain is created with $V_{tip}$= -10V. The gate lead is oriented perpendicular to the barrier region and is separated from the channel by 50 nm.

The gate-tuned drain current $I_D$ as a function of the source-drain voltage $V_{SD}$ (*I-V* characteristics) of the SketchFET was measured at various temperatures ranging from room temperature (295 K) down to 15 K. Representative curves for three gate biases ($V_{GD}$) and three temperatures are shown in Figure 1c. In general, positive gate biases increase the source-drain conductance while negative biases suppress it. One qualitative interpretation is that the gate electrode is shifting the bottom of conduction band at the barrier region through the Fermi level and thereby altering the carrier density as with a standard field-effect transistor. At room temperature, the channel conductance near zero source-drain bias can be tuned by the gate over more than four orders of magnitude. Most of the *I-V* curves are purely odd functions of $V_{SD}$, indicating that the field-tunable current flux is localized within the channel. At low temperatures and negative gate bias ($V_{GD}$=-2 V), an asymmetric *I-V* profile and slight negative differential resistance are observed (Fig. 1c(150K, 20K)), which is associated with the existence of leakage current from the gate lead[10].

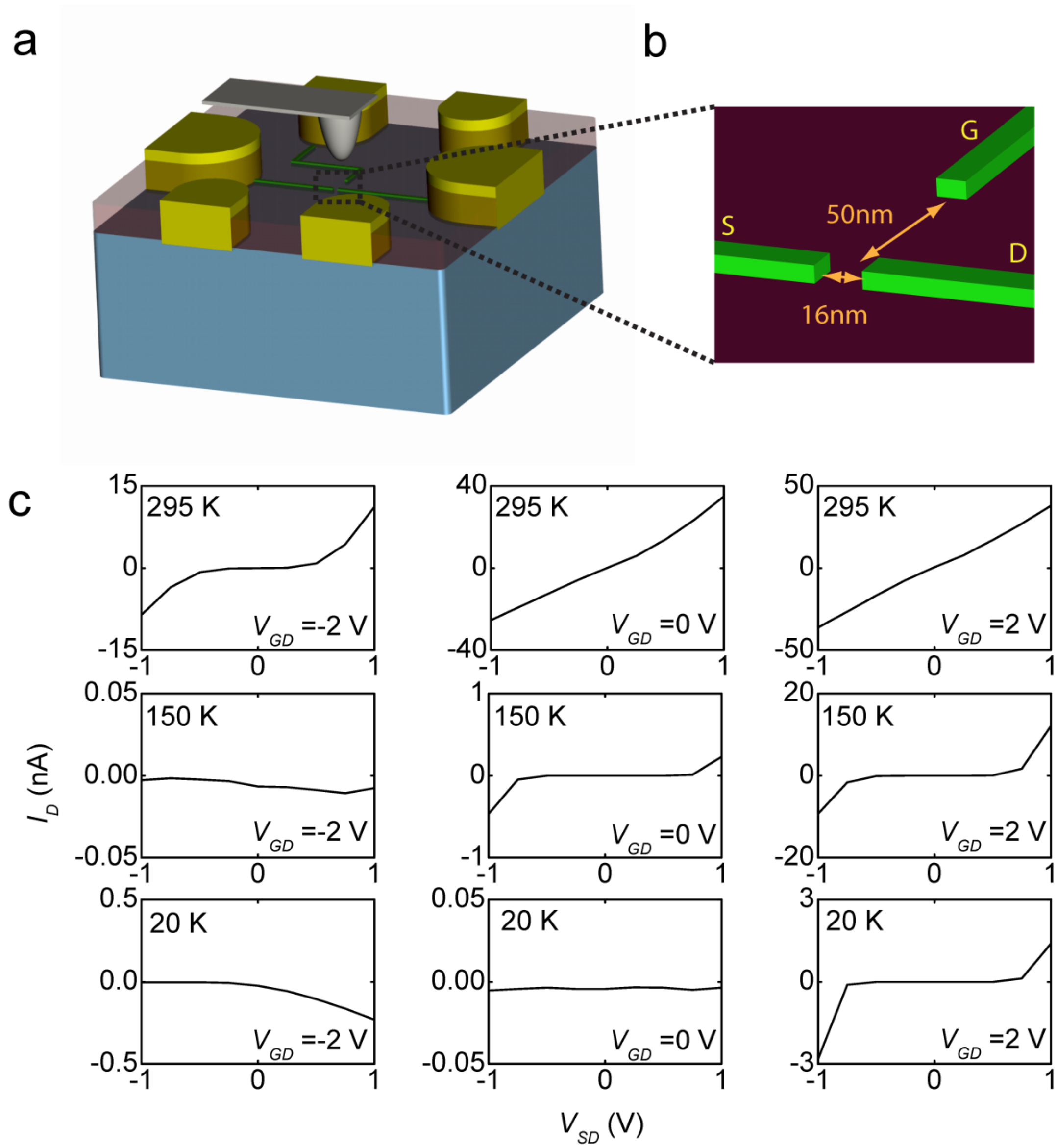

**Figure 1** (a) Illustration of a SketchFET structure written at the interface of 3uc-$LaAlO_3$/$SrTiO_3$ using a conductive AFM probe. (b) Illustration of the central junction area of the SketchFET structure used in the experiment. The SketchFET is composed of three 16 nm-wide nanowire sections: source (S), drain (D) and gate (G). The barrier width in the channel between the source and drain is 16 nm. The gate is separated from the channel by 50 nm. (c) *I*-*V* characteristic of the SketchFET channel between the source and drain from room temperature to cryogenic temperatures (295K, 150K, 20K) under different gating conditions $V_{GD}$ = -2 V, 0 V, 2 V.

Between room temperature and 150 K, the source-drain conductance decreases monotonically with decreasing temperature (Fig. 2a, b). Nanowires written without junctions typically exhibit a monotonically *increasing* conductance with decreasing temperature (Fig. S4a). Therefore, the potential barrier in the central junction of the SketchFET devices plays a dominant role in the decrease of the channel conductance. When the gate lead is also grounded, current measured at the drain can only flow from/to the source. Under these conditions, an unambiguous investigation of the barrier between the source and drain can be performed without the contribution of gate leakage current. Values of current flow into the drain when $V_{SD}$=2 V and current flow out of the drain when $V_{SD}$=-2 V are comparable (Fig. 2a), indicating a generally symmetric potential profile along the channel between the source and drain.

Arrhenius plots of the drain current $I_D$ as a function of temperature for various values of $V_{SD}$ and $V_{GD}$ are shown in Figure 2b. For a given $V_{GD}$, the temperature dependence of $I_D$ reflects a thermal activation characteristic. The saturation of $I_D$ at elevated temperatures can be accounted for by including the finite conductance of the source and drain nanowire leads which limits $I_D$ when the thermal activation rate is high. Numerical fits using a simple equivalent-circuit model (Fig. S3) produce good agreement with experimental results (Fig. 2b).

Figure 2c shows the extracted activation energies $E_a$ for different source and gate biases. When a non-zero gate voltage is applied, gate leakage may also contribute to the drain current; however, due to the much wider barrier between gate and channel (Fig. 1b), this leakage current is negligibly small above 150K. Therefore, the activation energies $E_a$ extracted provide a reasonably accurate quantification of the potential barrier between the source and drain. The results show that the source-drain energy barrier can be reduced either by increasing the gate bias or the magnitude of the source bias. Varying the gate and source bias between +/-2 V modulates the source-drain barrier by 0.5 eV, which is consistent with the $10^4$ on-off ratio of the SketchFET device. We note that higher ratios are in principle achievable if the nanowire leads conductance is reduced appropriately.

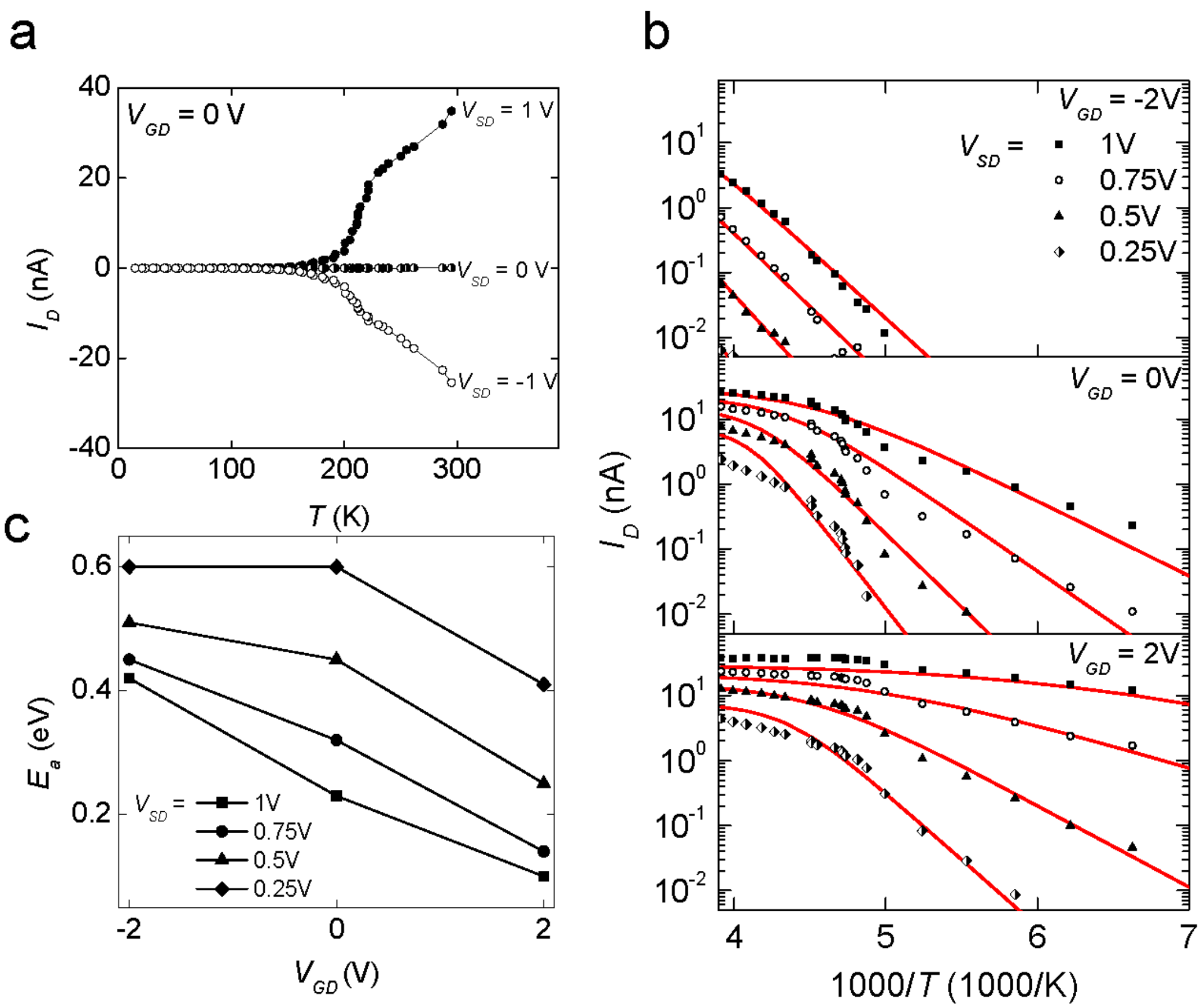

**Figure 2** (A) Source-drain current $I_D$ versus temperature for $V_{GD}$=0 V and $V_{SD}$ = 1 V, 0 V, -1 V. When the gate is grounded, drain current flows only from/to the source. (B) Arrhenius plot of drain current as a function of temperature at gate voltage $V_{GD}$ = -2 V, 0 V, 2 V, with source voltage $V_{SD}$ = 1 V, 0.75 V, 0.5 V, 0.25 V applied. Red lines represent fits to a transport model in which the thermally activated barrier between the source and drain is connected in series with nanowire leads of finite lead conductance. (C) Thermal activation energy $E_a$ extracted from data shown in (B).

When the source lead is grounded, only gate emitted/collected electrons contribute to the current measured at the drain. This leakage current is plotted as a function of temperature in Figure 3a at different gate biases. In the thermally activated region above 200 K, the magnitude of drain current $|I_D|$ is much larger for $V_{GD}>0$ than for $V_{GD}<0$. Activation energies $E_a^+ = 0.39$ eV and $E_a^- = 0.53$ eV are extracted from the Arrhenius plot for $V_{GD} = \pm 2$ V, respectively (Fig. 3b). The difference in the activation energy signifies an asymmetric shape of the barrier between gate and drain. Below 200 K, the thermally activated leakage current decays below our measurement limit. For the positive gate bias $V_{GD}$=2V, the flow of electrons from drain to gate remains small down to the lowest temperature measured. For negative gate bias $V_{GD}$=-2V, the flow of electrons from the gate exhibits two local maxima at $T_{C1}$=65 K and $T_{C2}$=25 K. We suggest that at low temperatures, the main contribution to the leakage current at $V_{GD}<0$ is quantum tunneling via Fowler-Nordheim (FN) field emission. As will be discussed in detail below, the FN field emission is highly sensitive to dielectric permittivity anomalies in the $SrTiO_3$ layer close to the interface with $LaAlO_3$. This sensitivity is directly responsible for the enhanced drain current at $T_{C1}$ and $T_{C2}$.

The electronic properties of 3uc-$LaAlO_3/SrTiO_3$ nanostructures are dominated by the smaller-bandgap $SrTiO_3$ top layers where the electrons are believed to be localized[2-7]. At room temperature, $LaAlO_3$ has a lattice constant of $a_{LAO}$ = 3.821 Å which is 2.2% smaller than the lattice constant of $SrTiO_3$, $a_{STO}$ = 3.905 Å. Epitaxial growth of coherently strained $LaAlO_3$ has a reciprocal effect on the near-surface $SrTiO_3$ layer, causing it to undergo multiple phase transitions at low temperatures[13,14]. These strain-induced near-surface structural phase transitions (and resulting distortions) have been observed using X-ray diffraction[15] and transmission electron microscopy[16].

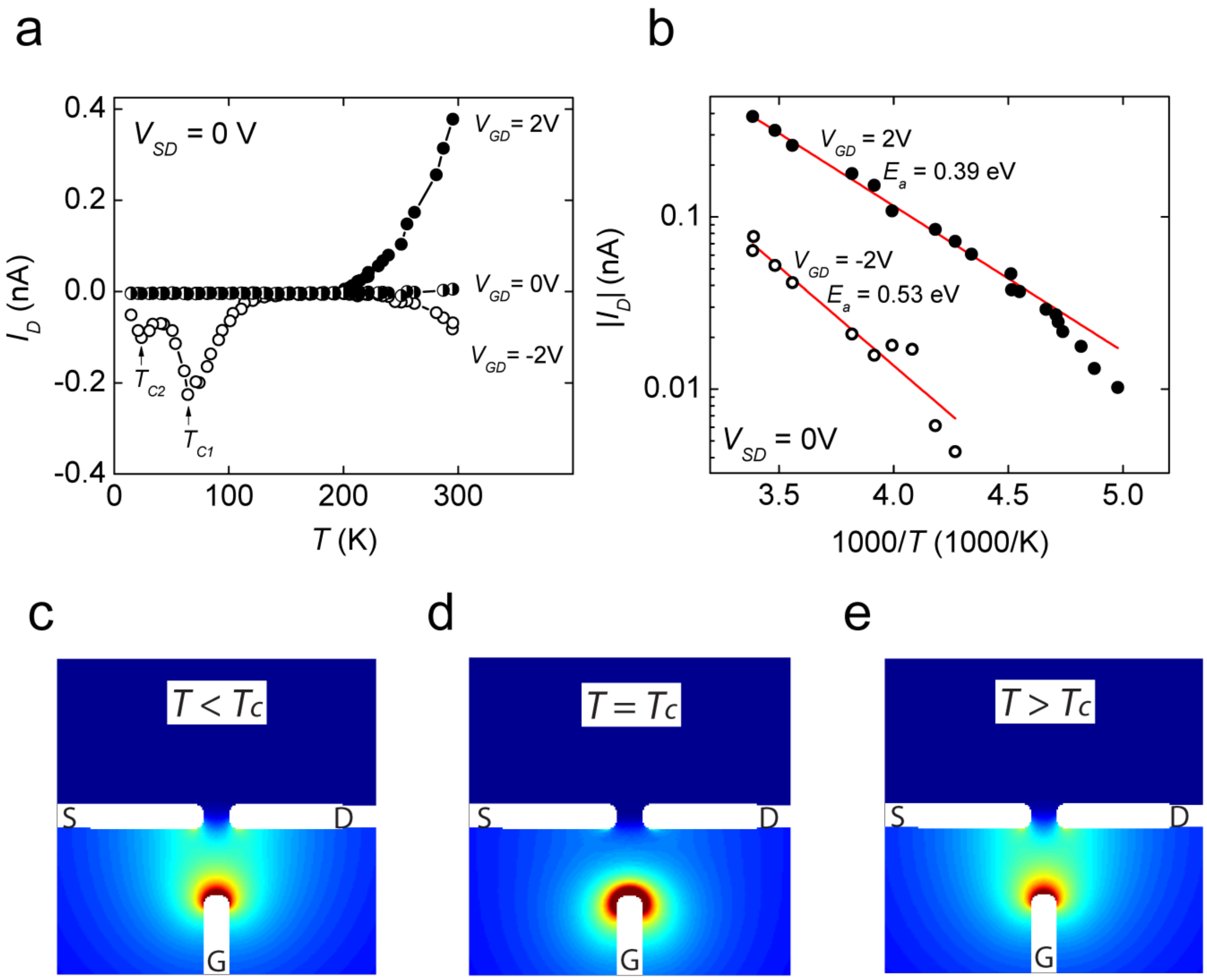

**Figure 3** Temperature dependence of gate leakage current. For $V_{SD}$=0 V, current only flows between the gate and drain. (a) Gate-drain current $I_D$ versus temperature for $V_{GD}$ = 2 V, 0 V, -2 V. Maxima in $|I_D|$ are observed at two temperatures $T_{C1}$=65 K and $T_{C2}$=25 K for $V_{GD}$=-2 V. (b) Arrhenius plot of leakage current above 200K at gate bias of 2V or -2V. Difference in the extracted thermal activation energies indicates an asymmetry in the potential profile between the gate and the channel. (c-e) Finite-element-method simulation of electric field strength in the central junction area of the SketchFET for $V_{GD}$=-2 V and temperatures below (C), at (D) and above (E) a structural phase transition. Near $T$=$T_C$, a marked increase in field strength at the gate electrode greatly raises the electron field emission rate.

$SrTiO_3$ is intrinsically a "high-k" dielectric. The relative dielectric permittivity $\varepsilon$ of unstrained bulk $SrTiO_3$ is approximately 300 at room temperature, and increases to ~20,000 or more at low temperatures as it approaches (but does not reach) a ferroelectric phase[17-20]. Both epitaxial strain[13,14,21] as well as applied electric fields[18,22-25] can strongly influence the dielectric properties, which in turn can affect the behavior of nanoscale devices.

Close to a structural phase transition, the dielectric permittivity $\varepsilon$ is highly sensitive to the local electric field strength. The temperature dependence of $\varepsilon$ is well described by Landau-Ginsburg-Devonshire (LGD) phenomenological models[26], capturing, for example, the well-known Curie-Weiss law behavior:

$$\frac{1}{\varepsilon}=\begin{cases}\dfrac{T-T_C}{C}, & T>T_C\\[2ex] 2\dfrac{T_C-T}{C}, & T<T_C\end{cases} \tag{1}$$

where $T_C$ is the Curie-Weiss temperature and $C$ is a constant. The electric-field dependence $\varepsilon(T,E)$ in $SrTiO_3$ is readily estimated within the simplest LGD model (see Supporting Information and Fig. S2). Using the calculated function $\varepsilon(T,E)$, the electric field strength over the 100 nm×100 nm SketchFET area is calculated self-consistently at different temperatures using a finite element method (Fig. 3c-e). At transition temperatures $T_{C1}$ (paraelectric to ferroelastic) and $T_{C2}$ (ferroelastic to ferroelectric), the diverging and highly field-sensitive dielectric permittivity greatly improves the field screening in the middle area between the gate, source and drain leads. As a consequence, the electric field $E$ is localized at the boundary of the gate lead with a significantly increased local intensity. The sharp increase in the electric field strength for negative $V_{GD}$ at $T \approx T_C$ greatly enhances the FN field emission rate $j \propto E^2 \exp\left(-\frac{b\Phi^{\frac{3}{2}}}{E}\right)$, where $\Phi$ is the work function of the emitter and $b$ is a constant[27]. For positive $V_{GD}$, the leakage current remains small because the drain lead, as the electron emitter, has significantly lower boundary field strength.

The low temperature maxima in leakage current at $V_{GD}$=-2 V vanish when $V_{SD}$ changes sign from 1 V to -1 V (Fig. 4a). This result can be explained by the decreased local field strength around the boundary of the gate lead (Fig. 4b-d). The sensitivity of the leakage current to the field strength near the emitter is a signature of FN field-emission processes. Dielectric anomalies also show up in source-drain current (Fig. 4e) but coexist with a residual tail of the thermally activated current. Varying the source voltage changes the local field strength along the electron-emitting source lead, which tunes the local maxima's intensity (Fig. 4f-h). Anomalies in conductance at 65 K and 25 K are also observed in nanowire structures where the potential barriers within the nanowires are formed unintentionally (Fig. S4b). One potentially important contribution not considered here is that the electron mobility within the nanowires may experience similar anomalies, thus increasing the attempt rate for electron tunneling through the barrier[28]. A full quantitative picture would need to take into account (self-consistently) the sharp variation in the dielectric permittivity within the nanowires themselves. Such detailed analysis extends beyond the scope of this paper.

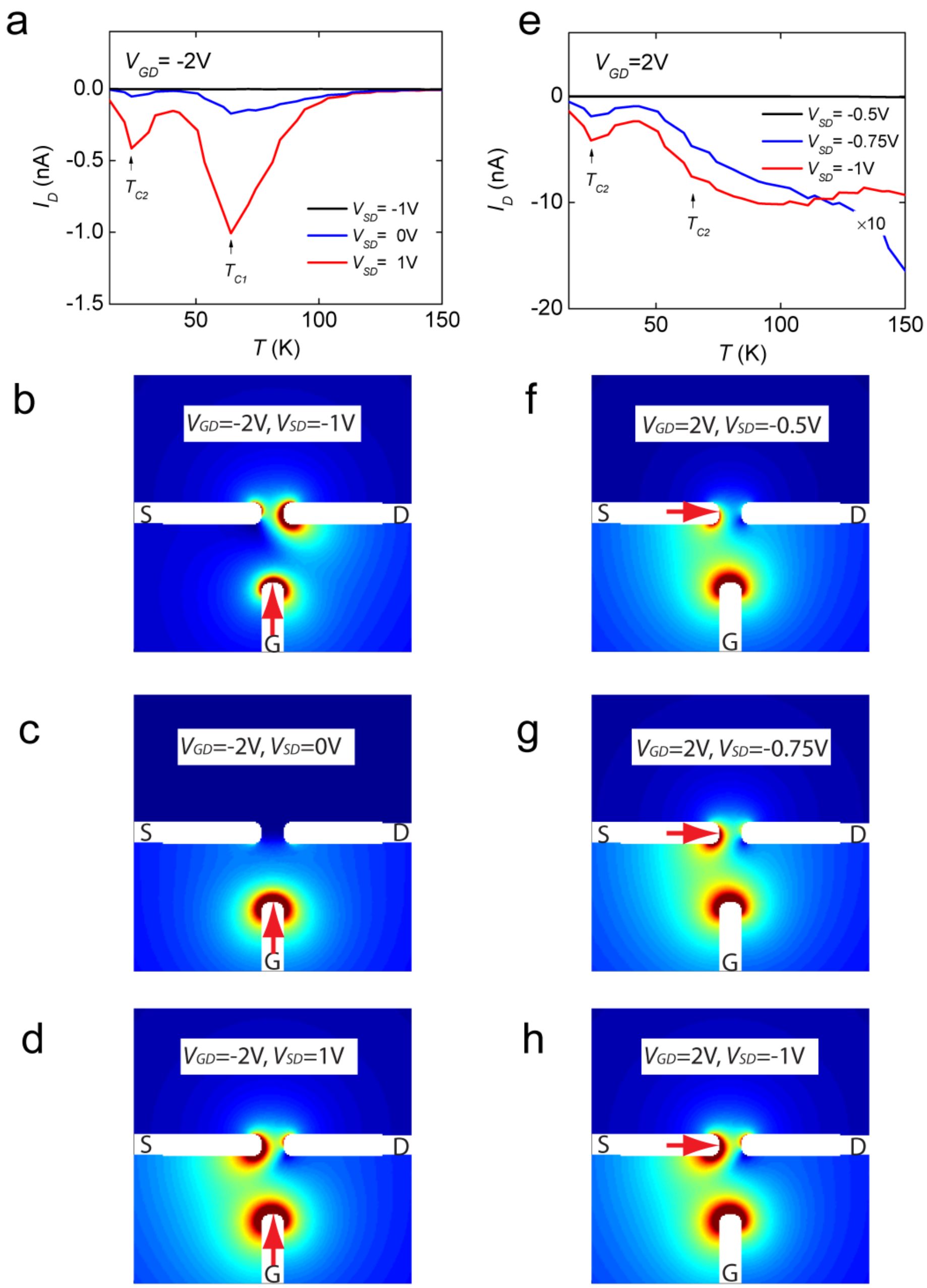

**Figure 4** (a) Measured drain current $I_D$ versus temperature for $V_{SD}$=-2 V and $V_{SD}$= 1 V, 0 V and -1 V. (b-d) Finite element method simulation of the electric field strength at the transition temperature in the central junction area of the SketchFET when -1V, 0V, 1V is applied to the source. Field strength at the boundary of the electron emitter (gate, annotated with red arrow) decreases as source bias increases, which reduces the electron emission rate. (e) Measured drain current $I_D$ versus temperature for $V_{GD}$=2 V and $V_{SD}$= -0.5 V, -0.75 V and -1 V. (f-h) Finite element method simulation of the electric field strength at the transition temperature in the central junction area of the SketchFET when -0.5 V, -0.75 V, -1 V is applied to the source. Field strength at the boundary of the electron emitter (source, annotated with red arrow) decreases as the absolute value of source bias decreases, which reduces the electron emission rate.

We have measured temperature-dependent transport in a nanoscale transistor (SketchFET), created at the 3uc-$LaAlO_3/SrTiO_3$ interface using a rewritable AFM lithography technique. The SketchFET maintains its transistor functionality down to the lowest temperatures measured, $T$=15 K. Between room temperature and 150 K, transport in SketchFET is dominated by thermal activation. Changing voltages applied to the gate and source electrodes within ±2V can tune the channel activation energy from 0.1 eV to 0.6 eV. Our experimental observation of four orders of magnitude on-off ratio is limited by the source-drain lead resistance. Below 150 K, Fowler-Nordheim quantum field emission dominates the transport. Sharp peaks of field emission current are observed at 65 K and 25 K, and are attributed to structural phase transitions in the $SrTiO_3$. This investigation marks the first step in characterizing energy landscape of oxide nanostructures, with implications for the performance of nanodevices at room temperature and at low temperatures.

**Acknowledgements**

We thank Stefan Thiel and Jochen Mannhart at the University of Augsburg for supplying the $LaAlO_3/SrTiO_3$ structures used for this work. We also thank Petro Maksymovych for helpful discussions. This work was supported by DARPA (W911NF-09-10258), ARO MURI (W911NF-08-1-0317) and NSF (DMR-0704022).

## References

1 Takagi, H. & Hwang, H. Y. An Emergent Change of Phase for Electronics. *Science* **327**, 1601 (2010).

2 Ohtomo, A. & Hwang, H. Y. A high-mobility electron gas at the $LaAlO_3/SrTiO_3$ heterointerface. *Nature* **427**, 423 (2004).

3 Thiel, S., Hammerl, G., Schmehl, A., Schneider, C. W. & Mannhart, J. Tunable quasi-two-dimensional electron gases in oxide heterostructures. *Science* **313**, 1942 (2006).

4 Basletic, M. *et al.* Mapping the spatial distribution of charge carriers in $LaAlO_3/SrTiO_3$ heterostructures. *Nature Materials* **7**, 621 (2008).

5 Huijben, M. *et al.* Electronically coupled complementary interfaces between perovskite band insulators. *Nature Materials* **5**, 556 (2006).

6 Pentcheva, R. & Pickett, W. E. Charge localization or itineracy at $LaAlO_3/SrTiO_3$ interfaces: Hole polarons, oxygen vacancies, and mobile electrons. *Phys. Rev. B* **74**, 7 (2006).

7 Janicka, K., Velev, J. P. & Tsymbal, E. Y. Quantum Nature of Two-Dimensional Electron Gas Confinement at $LaAlO_3/SrTiO_3$ Interfaces. *Phys. Rev. Lett.* **102**, 4 (2009).

8 Mannhart, J. & Schlom, D. G. Oxide Interfaces--An Opportunity for Electronics. *Science* **327**, 1607 (2010).

9 Cen, C. *et al.* Nanoscale control of an interfacial metal-insulator transition at room temperature. *Nature Materials* **7**, 298 (2008).

10 Cen, C., Thiel, S., Mannhart, J. & Levy, J. Oxide Nanoelectronics on Demand. *Science* **323**, 1026 (2009).

11 Xie, Y., Bell, C., Yajima, T., Hikita, Y. & Hwang, H. Y. Charge Writing at the $LaAlO_3/SrTiO_3$ Surface. *Nano Letters* **10**, 2588 (2010).

12 Bogorin, D. F. *et al.* Nanoscale rectification at the $LaAlO_3/SrTiO_3$ interface. *Appl. Phys. Lett.* **97**, 013102 (2010).

13 Haeni, J. H. *et al.* Room-temperature ferroelectricity in strained $SrTiO_3$. *Nature* **430**, 758 (2004).

14 Warusawithana, M. P. *et al.* A Ferroelectric Oxide Made Directly on Silicon. *Science* **324**, 367 (2009).

15 Vonk, V. *et al.* Interface structure of $SrTiO_3/LaAlO_3$ at elevated temperatures studied in situ by synchrotron x rays. *Phys. Rev. B* **75**, 5417 (2007).

16 Maurice, J. L. *et al.* Electronic conductivity and structural distortion at the interface between insulators $SrTiO_3$ and $LaAlO_3$. *Phys. Stat. Sol. A* **203**, 2209 (2006).

17 Barrett, J. H. Dielectric Constant in Perovskite Type Crystals. *Phys. Rev.* **86**, 118 (1952).

18 Fuchs, D., Schneider, C. W., Schneider, R. & Rietschel, H. High dielectric constant and tunability of epitaxial $SrTiO_3$ thin film capacitors. *J. Appl. Phys.* **85**, 7362 (1999).

19 Lippmaa, M. *et al.* Step-flow growth of $SrTiO_3$ thin films with a dielectric constant exceeding $10^4$. *Appl. Phys. Lett.* **74**, 3543 (1999).

20 Weaver, H. E. Dielectric properties of single crystals of $SrTiO_3$ at low temperatures. *Journal of Physics and Chemistry of Solids* **11**, 274 (1959).

21 Pertsev, N. A., Tagantsev, A. K. & Setter, N. Phase transitions and strain-induced ferroelectricity in $SrTiO_3$ epitaxial thin films. *Phys. Rev. B* **61**, R825 (2000).

22 Antons, A., Neaton, J. B., Rabe, K. M. & Vanderbilt, D. Tunability of the dielectric response of epitaxially strained $SrTiO_3$ from first principles. *Phys. Rev. B* **71**, 024102 (2005).

23 Bouzehouane, K. *et al.* Enhanced dielectric properties of $SrTiO_3$ epitaxial thin film for tunable microwave devices. *Appl. Phys. Lett.* **80**, 109 (2002).

24 Hemberger, J., Lunkenheimer, P., Viana, R., Bohmer, R. & Loidl, A. Electric-Field-Dependent Dielectric-Constant And Nonlinear Susceptbility in $SrTiO_3$. *Phys. Rev. B* **52**, 13159 (1995).
25 Park, K. C. & Cho, J. H. Electric field dependence of ferroelectric phase transition in epitaxial $SrTiO_3$ films on $SrRuO_3$ and $La_{0.5}Sr_{0.5}CoO_3$. *Appl. Phys. Lett.* **77**, 435 (2000).
26 Lines, M. E. & Glass, A. M. *Principles and Applications of Ferroelectrics and Related Materials*. (Oxford University Press, 1996).
27 Fowler, R. H. & Nordheim, L. Electron Emission in Intense Electric Fields. *Proceedings of the Royal Society of London. Series A* **119**, 173 (1928).
28 Jena, D. & Konar, A. Enhancement of carrier mobility in semiconductor nanostructures by dielectric engineering. *Phys. Rev. Lett.* **98**, 136805 (2007).

## Supplemental Information

# Thermal Activation and Quantum Field Emission in a Sketch-Based Oxide Nano Transistor

Cheng Cen[1], Daniela F. Bogorin[1], Jeremy Levy[1*]

[*]Corresponding author: jlevy@pitt.edu

[1]Department of Physics and Astronomy, University of Pittsburgh, Pittsburgh, PA 15260, USA

### Measurement of Nanowire Width

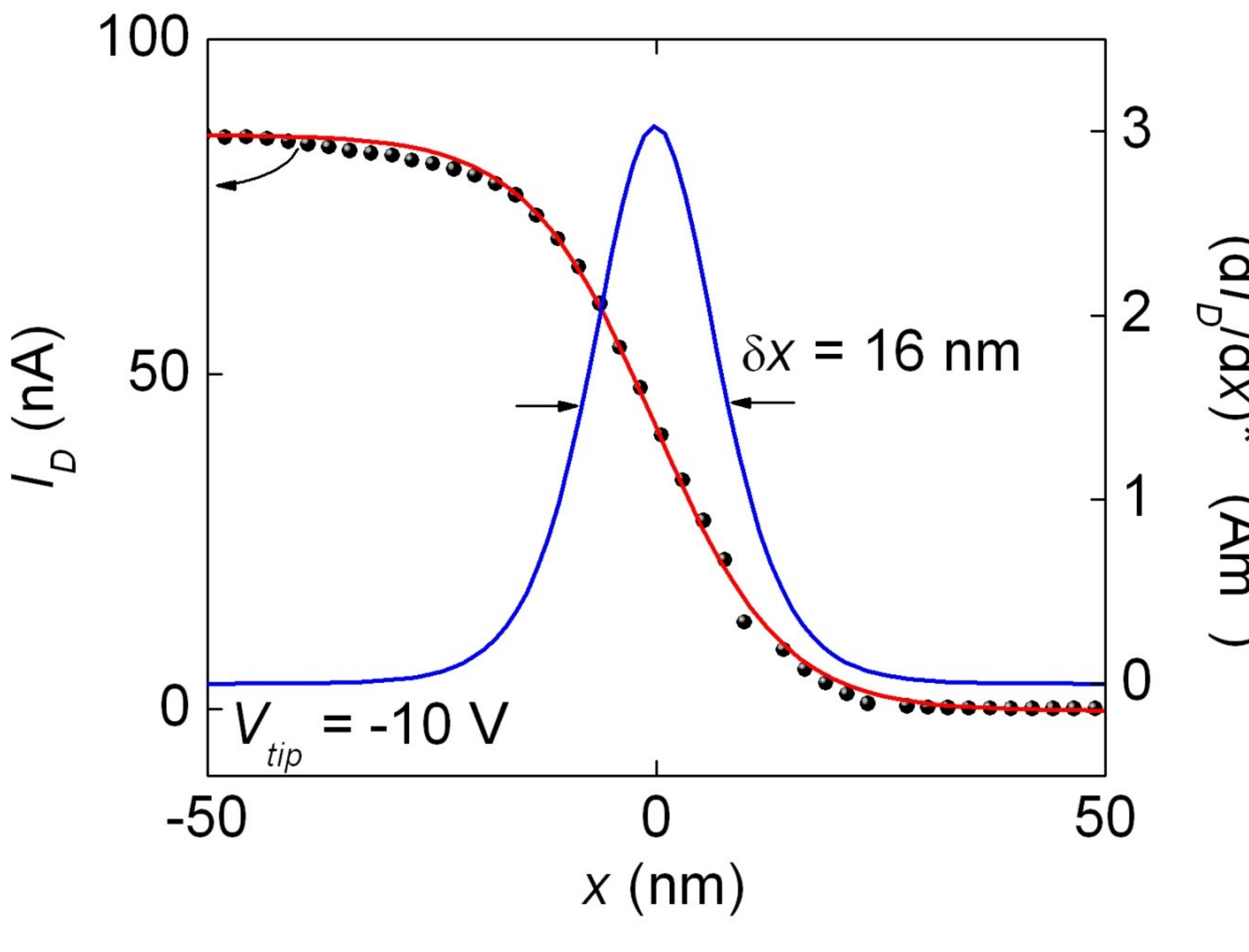

**Figure S1.** For the SketchFET structure, the source-drain current is measured as a function of the tip position $x$ across the wire, while cutting the wire with the tip biased negatively. A sharp drop in conductance occurs when the tip position crosses the wire. The decrease in conductance can be fit to a profile $I(x)=I_0-I_1\tanh(x/h)$ . Also plotted is the deconvolved differential current $(dI/dx)^{*-1}$. Cutting was done with -10 V tip bias. The deconvolved differential current shows a full width at half maximum of $\delta x$ = 16 nm.

## Landau-Ginzburg-Devonshire (LGD) Theory

LGD theory provides a successful phenomenology description of many perovskite ferroelectrics. Based on this theory, the free energy of a ferroelectric material can be expressed as[1]:

$$G = \left(\frac{\alpha}{2}\right)P^2 + \left(\frac{\gamma}{4}\right)P^4 + \left(\frac{\delta}{6}\right)P^6 - EP$$

where $E$ is the electric filed intensity and $P$ is the polarization density. Close to a second order ferroelectric phase transition, $\gamma$ and $\delta$ are temperature-independent positive constants, $\alpha = \beta(T - T_C)$, and $\beta$ is a positive constant. Minimization of the free energy leads to:

$$E = \beta(T - T_C)P + \gamma P^3 + \delta P^5,$$

from which the reciprocal susceptibility can be derived:

$$\kappa = \frac{\partial E}{\partial P} = \beta(T - T_C) + 3\gamma P^2 + 5\delta P^4.$$

The dielectric permittivity is given by:

$$\varepsilon = 1 + \kappa^{-1} = 1 + (\beta(T - T_C) + 3\gamma P^2 + 5\delta P^4)^{-1}.$$

Close to the phase transition temperature $T_C$, $\varepsilon$ diverges and is highly field-sensitive. Figure S2 shows $\varepsilon(E, T)$ using $\beta = 1.6 \times 10^5, \gamma = \delta = 1, T_C = 25$ K.

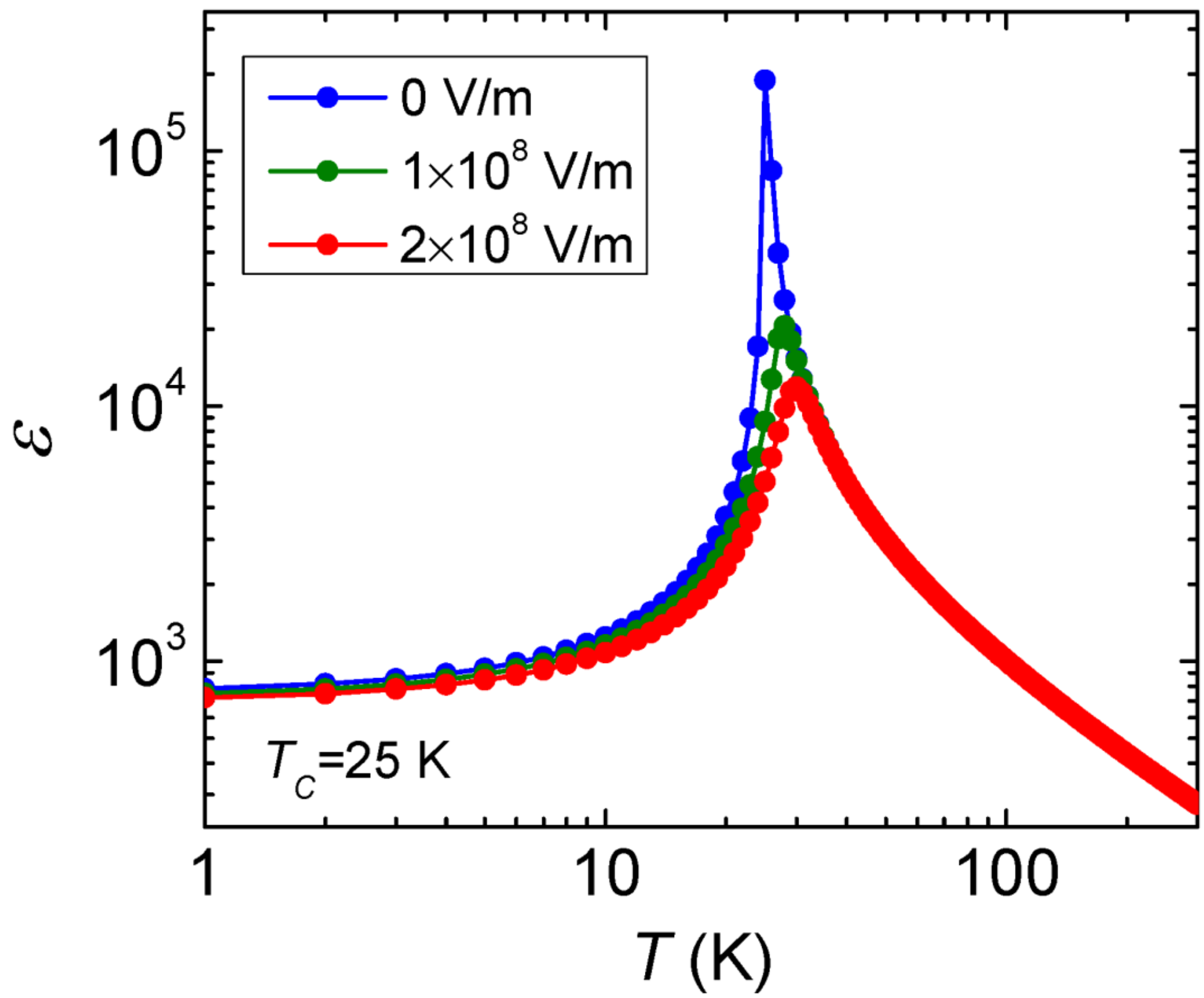

**Figure S2.** Simulation of the dielectric permittivity $\varepsilon(E,T)$ of $SrTiO_3$ during a structural phase transition according to the LGD model. Approaching the transition temperature $T_C$, $\varepsilon$ increases dramatically and also becomes highly sensitive to the local electric field strength.

**Finite Nanowire Lead Conductance**

Close to room temperature, the thermal activation rate is high in the central junction, and the finite conductance of the source and drain nanowire leads need to be considered. Figure S3 shows a simple circuit model that incorporates both the transistor as well as the source and drain lead resistances.

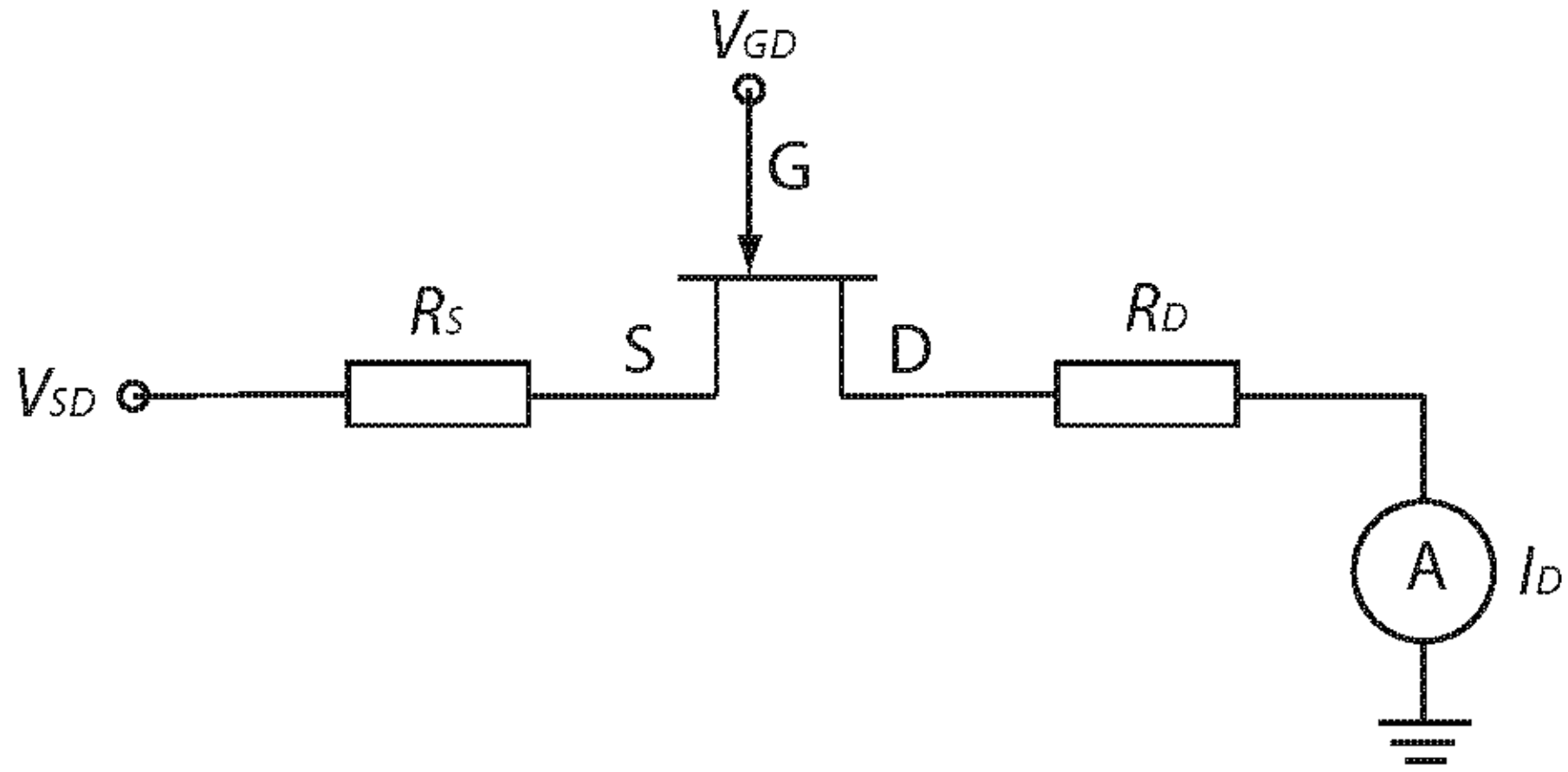

**Figure S3** Equivalent circuit model in which the source and drain nanowire leads are modeled as resistors ($R_S$, $R_D$) connected in series with the central junction of the SketchFET.

The drain current can be expressed as:

$$I_D = V_{SD} \frac{\frac{1}{R_S + R_D} \times G_0 \exp\left(-\frac{E_a}{kT}\right)}{\frac{1}{R_S + R_D} + G_0 \exp\left(-\frac{E_a}{kT}\right)}$$

where $G_0 \exp\left(-\frac{E_a}{kT}\right)$ is the thermally activated conductance of the central junction. In all the fitting data presented in Figure 2B, a single value of $\frac{1}{R_S+R_D} = 3 \times 10^{-8}$ S is used.

**Temperature Dependence Conductance of Nanowires**

In most instances, the conductance of a nanowire written by AFM lithography increases with decreasing temperature, evidence for metallic behavior. An example of such a response is shown in Figure S4A. However, in some cases there may be unintentionally generated potential barriers along the wire. In such cases (for example Figure S4B), the conductance is found to decrease with decreasing temperature. We observe conductance anomalies in these "naturally" occurring potential barriers at temperatures $T_{C1}$ and $T_{C2}$ that are consistent with the artificial barriers described in the main text.

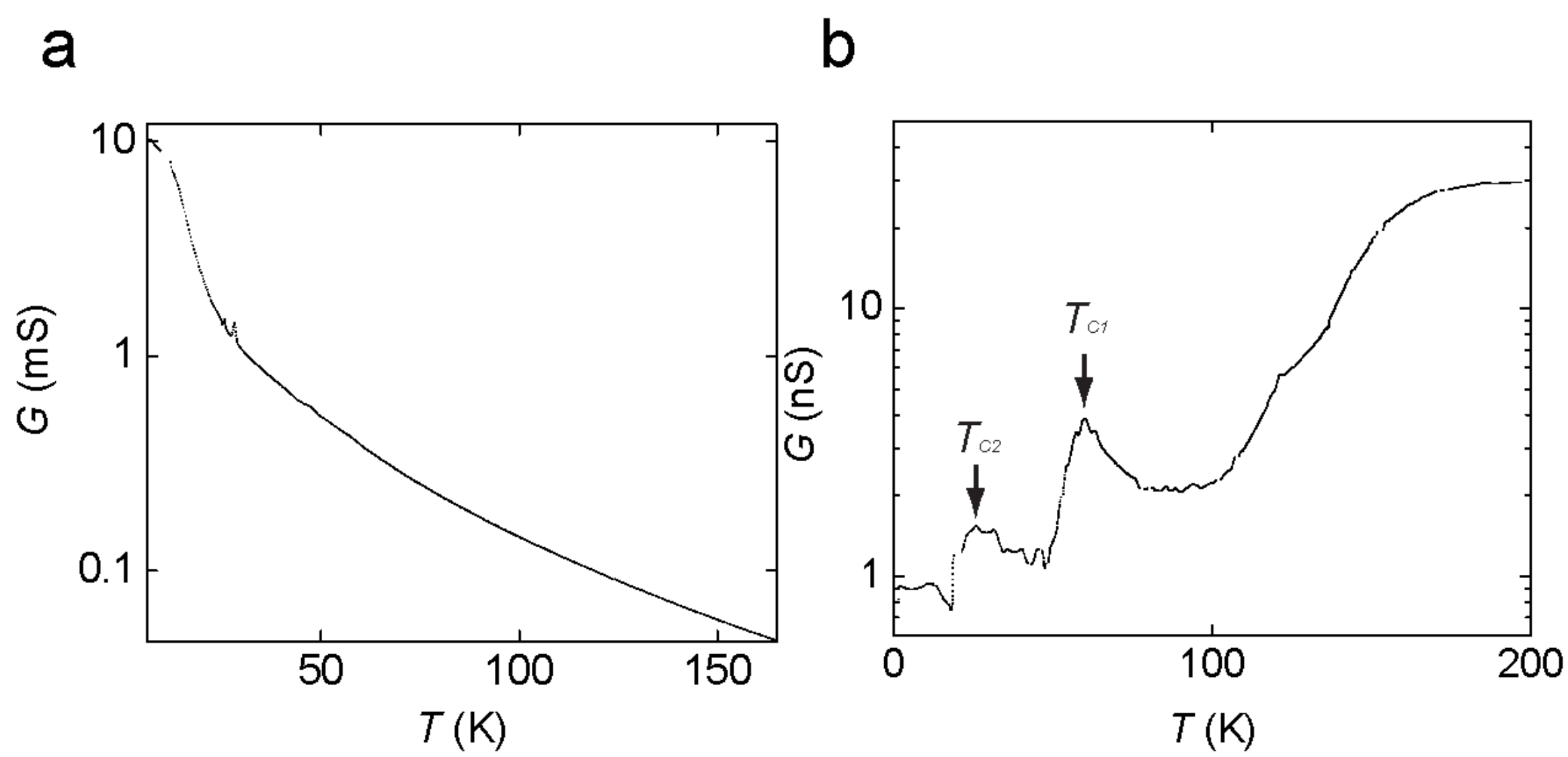

**Figure S4.** (a) Example of a highly conducting nanowire whose conductance increases monotonically with decreasing temperature. (b) Example of a weakly conducting nanowire whose conductance decreases non-monotonically with decreasing temperature. Observed anomalies at $T_{C1}$ and $T_{C2}$ are consistent with those measured in the SketchFET device.

## Reference

1. Lines, M. E.; Glass, A. M., *Principles and Applications of Ferroelectrics and Related Materials*. Oxford University Press: 1996.